# Cooper-pair based photon entanglement without isolated emitters.

Alex Hayat[1], Hae-Young Kee[1,2], Kenneth S. Burch[1,2,3] and Aephraim M. Steinberg[1,3]

[1]*Department of Physics, University of Toronto, Toronto, Ontario, Canada M5S 1A7*
[2]*Centre for Quantum Materials, University of Toronto, 60 St. George Street, Toronto, Ontario, Canada M5S 1A7*
[3]*Centre for Quantum Information and Quantum Control and the and Institute for Optical Sciences, University of Toronto, 60 St. George Street, Toronto, Ontario, Canada M5S 1A7*

We propose an efficient approach for generation of entangled photons, based on Cooper-pair luminescence in semiconductors, which does not require isolated emitters such as single atoms or quantum dots. We show that in bulk materials, electron-spin entanglement in Cooper pairs should not be expected to be translated into pure entangled photons despite the selection rules, due to mixing introduced by light-hole – heavy-hole degeneracy. Semiconductor quantum wells, however, remove this degeneracy, allowing efficient photon entanglement generation in simple electrically-driven structures, taking advantage of the superconducting macroscopic coherence. The second-order decay of two-electron states in Cooper-pair luminescence leaves no which-path information, resulting in perfect coherence between two pathways and hence in principle, perfect entanglement. We calculate the purity of the entangled-photon state and find that it increases for larger light-hole – heavy-hole energy splitting and for lower temperatures. These results provide insights into light-matter interaction in solids and enable realization of quantum photonics based on matter condensates.

Entanglement is among the most intriguing aspects of quantum mechanics, contradicting the local realism of classical physics via the violation of Bell inequalities [1, 2]. Moreover, various applications in the growing field of quantum information science such as quantum cryptography [3], computing [4], and metrology [5] require efficient sources of entangled photon pairs. The most widely used current technique of generating photon pairs – parametric downconversion [2,6] – is limited by the weakness of the non-resonant $\chi^{(2)}$ nonlinearity [7], and requires phasematching and optical excitation, which prevent its integration into compact photonic circuits. A recently observed process of semiconductor two-photon emission [8,9] enables compact electrically-pumped photon-pair sources with nanophotonic enhancement [10]; however the efficiency of such sources is relatively low due to the non-resonant second-order transition. Cascaded emission from biexcitons in semiconductor quantum dots (QD) allows generation of entanglement in miniature devices, where each discrete QD energy level can only be occupied by two fermions with opposite spins [11]. Anisotropic exchange splitting in QDs, however, leaves which-path information and thus significantly complicates entanglement generation [12]. Hybrid devices based on semiconductor-superconductor structures are a rapidly growing field [13,14,15], including QDs and nanocrystals integrated into Josephson light-emitting diodes [16,17]. These hybrid devices were proposed as enhanced QD entanglement sources [18, 19, 20] based on opposite-spin electrons at each discrete energy level. However, these isolated emitters have inherently low emission rates, and require sophisticated fabrication methods, carrier injection and light extraction techniques. Moreover, although superconductivity can enhance emission rates, it is not necessary for the generation of entanglement in isolated emitters such as QDs where discrete levels allow entanglement generation without superconductivity. [11,12]

Two-dimensional and bulk semiconductor structures, at the core of the existing semiconductor optoelectronic infrastructure, are significantly simpler and more efficient, and they have been shown lately to result in enhanced electrically-driven light emission when combined with superconductors [21,22,23]. In contrast to QD-based sources, however, the continuum of states in these structures allows population of infinitesimally close states by electrons with the same spin – preventing any polarization correlation between the emitted photons without superconductivity.

Here we show that Cooper-pair electron-spin entanglement provides a unique source of entangled photon pairs in hybrid superconductor-semiconductor quantum well (QW) structures, which cannot be realized without superconductivity. This approach takes advantage of the macroscopic coherence of the superconducting state for enhanced entanglement generation rates in a relatively simple electrically-pumped structure. Despite the fact that electrons in Cooper pairs are entangled, our analysis shows that a bulk semiconductor-based emitter with spin-orbit coupled angular momentum states results in a mixed photon state (not a pure state, and thus not maximally entangled) due to the contribution of both light hole (LH) and heavy hole (HH) bands. In a two-dimensional quantum well, on the other hand, pure entangled states are produced due to the lifting of LH/HH degeneracy, and angular momentum selection rules. The recombination of a Cooper pair is a second-order transition [23] that leaves no which-path information in the final state, in contrast to a pair of single-particle first-order transitions. We explicitly calculate the density matrices of the emitted photon-pair states. Small LH/HH splitting and high temperatures can introduce some mixing into the emitted states; nevertheless, for temperatures sufficiently low to maintain superconductivity and for typical QW dimensions, our calculations predict essentially pure entangled photon states.

In the presence of the proximity effect, whereby a superconducting state is induced in a semiconductor, the conduction electrons are not independent, but rather form a many-body Bardeen-Cooper-Schrieffer (BCS) [24] state yielding macroscopic coherence and enhanced emission [21]. Furthermore, the second-order emission in a BCS state is a two-electron transition resulting in photon-pair emission, in contrast to the single-electron transitions in usual nonlinear optics [2,6-10]. This two-electron second-order process in a coherent BCS state is what enables both pure entangled state generation and high emission rates in the hybrid semiconductor-superconductor QW structure.

We consider a superconducting proximity region induced in a direct-bandgap semiconductor [21, 25], where the superconducting gap $2\Delta$ is in the semiconductor conduction band (CB) with electrons in a BCS state, while the valence band is in the normal state of holes (Fig. 1 a). Near the Brillouin zone center, the CB electrons [26] have total spin-orbit coupled angular momentum $J_z = \pm 1/2$. Electrons in Cooper pairs in an $s$-wave superconductor are in an angular momentum singlet state [24]:

$$|\Psi\rangle = \frac{1}{\sqrt{2}}\left(|\uparrow\rangle_1|\downarrow\rangle_2 - |\downarrow\rangle_1|\uparrow\rangle_2\right) \qquad (1)$$

where the $|\uparrow\rangle$ and $|\downarrow\rangle$ denote electron states with $J_z = 1/2$ and $J_z = -1/2$ respectively. Recombination of such singlet states in a semiconductor with normal holes, however, does not necessarily result in the emission of entangled photon pairs. The two valence bands with significant populations of holes in typical direct-bandgap semiconductors – the LH band with $J_z = \pm 1/2$ and the HH band with $J_z = \pm 3/2$, are degenerate at zero crystal momentum. The selection rules for recombination of a HH and a CB electron allow transitions only with angular momentum change of $\Delta J_z = \pm 1$ [26]. For a singlet CB electron Cooper-pair state, such transitions should result in polarization-correlated photon pairs. However, the presence of the second energy-degenerate LH band with additional allowed transitions degrades polarization correlations (Fig. 1 a).

In our proposed scheme, the superconducting state is induced in a semiconductor QW, where the LH/HH degeneracy is lifted (Fig. 1 b). The two-dimensional QW structures enable injection of a very large number of Cooper pairs, resulting in carrier densities comparable to bulk materials – in contrast to the isolated zero-dimensional QDs. The macroscopic coherence of the BCS state can enhance the emission rate even further [21]. The CB-HH recombination emission results in polarization-correlated photons – however, for polarization *entanglement* to result, there must also be no which-path information in the final state. In a naïve single-particle description of Cooper-pair recombination, the final state might appear to maintain information on the recombination paths of each electron-hole pair, where the angular momentum of each electron-hole pair appears to be translated to a well-defined photon polarization (Fig. 1 b). However this picture, which describes Cooper-pair recombination as two separate first-order transitions, is incorrect, because no single-electron states exist inside the superconducting gap at the Fermi level. An electron Cooper pair, therefore, must recombine with a pair of holes in one second-order transition via a virtual state (Fig. 1 c). We show that this second-order transition preserves only the total angular momentum, whereas the polarization of each individual photon is not defined. Thus polarization-correlated photons with undefined individual polarization can be entangled. The photons in the pair are tagged by their colors (wavenumbers), $q_\mu$ and $q_\nu$, selected by spectral filtering, and are emitted in the direction of growth of the QW.

In addition to the entangled-photon pair emission, one-photon emission will also occur in the proposed device at an energy given by the bandgap $E_{BG} = \left(E_{CB} - E_{HH}\right)$. The one-photon emission will not result in entangled-photon pairs; however, it can be easily separated from the entangled-photon pair emission using spectral filtering of $q_\mu$ and $q_\nu$. The entangled-photon emission in Cooper-pair recombination is a second-order process occurring via a virtual state (Fig. 1 c), and not a cascade of one-photon emission events (Fig. 1b). Therefore, only the sum of the energies in the photon pair is fixed by the band gap $q_\mu + q_\nu = 2E_{BG}$, but this does not determine the energy of each individual photon to be $E_{BG}$. In a second-order process, the two emitted photons can be at energies very different from that of the one-photon emission $q_\mu = E_{BG} + \delta E$ and $q_\nu = E_{BG} - \delta E$, as long as the total energy is conserved $q_\mu + q_\nu = 2E_{BG}$ [2,6-10]. Therefore, for photons selected at energies $q_\mu = E_{BG} + \delta E$ and $q_\nu = E_{BG} - \delta E$, different from $E_{BG}$ by more than the thermal width $kT$, there is no corresponding one-photon emission and thus the entangled photon state will not be affected. We calculate the emitted state explicitly for sufficiently large HH-LH splitting yielding a pure polarization-entangled state.

In contrast to previous calculations [23] our model must include the entire spin-orbit coupled angular momentum $J$ in the interaction and the polarization $\sigma$ of the photons. Therefore, the light-matter coupling Hamiltonian in the interaction picture with $\hbar = c = 1$ is:

$$H_I = \sum_{\mathbf{k},\mathbf{q},\sigma,J} \left( B_{\mathbf{k},\mathbf{q}} b_{-(\mathbf{k}-\mathbf{q}),-J} c_{\mathbf{k},J+\sigma}^\dagger a_{\mathbf{q},\sigma}^\dagger + B_{\mathbf{k},\mathbf{q}}^* b_{-(\mathbf{k}-\mathbf{q}),-J}^\dagger c_{\mathbf{k},J+\sigma}^\dagger a_{\mathbf{q},\sigma} \right) \qquad (2)$$

where $B_{\mathbf{k},\mathbf{q}}$ is the coupling energy, $b_{\mathbf{k},J}^\dagger$ and $c_{\mathbf{k},J}^\dagger$ hole and electron creation operators with crystal momentum **k** and angular momentum $J$, and $a_{\mathbf{q},\sigma}^\dagger$ is the photon creation operator with linear momentum **q** and polarization $\sigma$. The initial state is given by $|\chi_0\rangle = |0\rangle |FS\rangle |BCS\rangle$, where $|0\rangle$ represents the vacuum of the photon field, $|FS\rangle$ denotes the Fermi sea of holes, and $|BCS\rangle$ is the electron superconducting BCS state. The hole thermal distribution is accounted for by integration over the Fermi-Dirac distribution. The second-order contribution to the final state is given by

$$|\chi_t\rangle = \int_{-\infty}^{t} dt_1 \int_{-\infty}^{t_1} dt_2 H_I(t_1) H_I(t_2) |\chi_0\rangle \quad , \quad \text{where} \quad H_I(t) = e^{iH_0 t} H_I e^{-iH_0 t} \quad \text{and}$$

$$H_0 = \sum_{\mathbf{q},\sigma} \omega_q^{ph} a_{\mathbf{q},\sigma}^\dagger a_{\mathbf{q},\sigma} + \sum_{\mathbf{k},J} \omega_k c_{\mathbf{k},J}^\dagger c_{\mathbf{k},J} + \sum_{\mathbf{k}',J'} \omega_{k'}' b_{\mathbf{k}',J'}^\dagger b_{\mathbf{k}',J'}.$$ This interaction can be described by

two-vertex Feynman diagrams (Fig. 2 a). The double-arrowed electron propagators describe the Green functions resulting from non-vanishing $\langle BCS | c_{\mathbf{k}_1,J_1}^\dagger c_{\mathbf{k}_2,J_2}^\dagger | BCS \rangle$ terms in the superconducting state [27]. This Green function permits pair emission through a single connected second-order Feynman diagram, in contrast to the disconnected pair of first-order single-electron transitions. It is important to note that the initial electron state is not a single Cooper pair in a singlet state, but rather a many-body BCS state. Therefore the entangled photon state emitted from the second-order transition does not have to be in a singlet state. The color-specific two-photon polarization state is fully described by the density matrix $\rho(q_\mu, q_\nu)$, whose elements are given by the following expectation value in the final state [12]:

$$\rho(q_\mu, q_\nu)_{\sigma_\alpha, \sigma_\beta, \sigma_\gamma, \sigma_\delta} = \langle \chi_t | a_{\mathbf{q}_\mu,\sigma_\alpha}^\dagger a_{\mathbf{q}_\nu,\sigma_\beta}^\dagger a_{\mathbf{q}_\mu,\sigma_\gamma} a_{\mathbf{q}_\nu,\sigma_\delta} | \chi_t \rangle =$$

$$= \int_{-\infty}^{t} dt_1 \int_{-\infty}^{t_1} dt_2 \int_{-\infty}^{t} dt_3 \int_{-\infty}^{t_3} dt_4 \langle \chi_0 | H_I(t_1) H_I(t_2) a_{\mathbf{q}_\mu,\sigma_\alpha}^\dagger a_{\mathbf{q}_\nu,\sigma_\beta}^\dagger a_{\mathbf{q}_\mu,\sigma_\gamma} a_{\mathbf{q}_\nu,\sigma_\delta} H_I(t_3) H_I(t_4) | \chi_0 \rangle \quad (3)$$

This calculation can be described by two kinds of Feynman diagrams (Fig. 2 b), which can be used to obtain the qualitative structure of the density matrix conveniently; however, a straightforward integration enables the quantitative calculation of the amplitudes. The integrand in Eq. 3 can be split into the photon $(I_{ph})$, electron $(I_e)$, and hole $(I_h)$ terms:

$$\langle\chi_0|H_I(t_1)H_I(t_2)a^\dagger_{\mathbf{q}_\mu,\sigma_\alpha}a^\dagger_{\mathbf{q}_\nu,\sigma_\beta}a_{\mathbf{q}_\mu,\sigma_\gamma}a_{\mathbf{q}_\nu,\sigma_\delta}H_I(t_3)H_I(t_4)|\chi_0\rangle =$$
$$= \sum_{\mathbf{k}_1..\mathbf{k}_4,\mathbf{q}_1..\mathbf{q}_4,\sigma_1..\sigma_4,J_1..J_4} B^*_{\mathbf{k}_1,\mathbf{q}_1}B^*_{\mathbf{k}_2,\mathbf{q}_2}B_{\mathbf{k}_3,\mathbf{q}_3}B_{\mathbf{k}_4,\mathbf{q}_4}I_{ph}I_h I_e \times \qquad (4)$$
$$\times e^{-i(\omega^{ph}_{\mathbf{q}_1}-\omega_{\mathbf{k}_1-\mathbf{q}_1})t_1}e^{-i(\omega^{ph}_{\mathbf{q}_2}-\omega_{\mathbf{k}_2-\mathbf{q}_2})t_2}e^{i(\omega^{ph}_{\mathbf{q}_3}-\omega_{\mathbf{k}_3-\mathbf{q}_3})t_3}e^{i(\omega^{ph}_{\mathbf{q}_4}-\omega_{\mathbf{k}_4-\mathbf{q}_4})t_4}$$

The photon term is calculated to be:

$$I_{ph} = (\delta_{\sigma_1,\sigma_\beta}\delta_{\sigma_2,\sigma_\alpha}\delta_{\sigma_3,\sigma_\delta}\delta_{\sigma_4,\sigma_\gamma}\delta_{\mathbf{q}_1,\mathbf{q}_\nu}\delta_{\mathbf{q}_2,\mathbf{q}_\mu}\delta_{\mathbf{q}_3,\mathbf{q}_\nu}\delta_{\mathbf{q}_4,\mathbf{q}_\mu} +$$
$$+\delta_{\sigma_1,\sigma_\alpha}\delta_{\sigma_2,\sigma_\beta}\delta_{\sigma_3,\sigma_\delta}\delta_{\sigma_4,\sigma_\gamma}\delta_{\mathbf{q}_1,\mathbf{q}_\mu}\delta_{\mathbf{q}_2,\mathbf{q}_\nu}\delta_{\mathbf{q}_3,\mathbf{q}_\nu}\delta_{\mathbf{q}_4,\mathbf{q}_\mu} +$$
$$+\delta_{\sigma_1,\sigma_\beta}\delta_{\sigma_2,\sigma_\alpha}\delta_{\sigma_3,\sigma_\gamma}\delta_{\sigma_4,\sigma_\delta}\delta_{\mathbf{q}_1,\mathbf{q}_\nu}\delta_{\mathbf{q}_2,\mathbf{q}_\mu}\delta_{\mathbf{q}_3,\mathbf{q}_\mu}\delta_{\mathbf{q}_4,\mathbf{q}_\nu} +$$
$$+\delta_{\sigma_1,\sigma_\alpha}\delta_{\sigma_2,\sigma_\beta}\delta_{\sigma_3,\sigma_\gamma}\delta_{\sigma_4,\sigma_\delta}\delta_{\mathbf{q}_1,\mathbf{q}_\mu}\delta_{\mathbf{q}_2,\mathbf{q}_\nu}\delta_{\mathbf{q}_3,\mathbf{q}_\mu}\delta_{\mathbf{q}_4,\mathbf{q}_\nu}) \qquad (5)$$

and the hole term is:

$$I_h = f^p_{\mathbf{k}_i}f^p_{\mathbf{k}_j}\left(\delta_{\mathbf{k}_1-\mathbf{q}_1,\mathbf{k}_3-\mathbf{q}_3}\delta_{J_1,J_3}\delta_{\mathbf{k}_2-\mathbf{q}_2,\mathbf{k}_4-\mathbf{q}_4}\delta_{J_2,J_4} - \delta_{\mathbf{k}_1-\mathbf{q}_1,\mathbf{k}_4-\mathbf{q}_4}\delta_{J_1,J_4}\delta_{\mathbf{k}_2-\mathbf{q}_2,\mathbf{k}_3-\mathbf{q}_3}\delta_{J_2,J_3}\right) \qquad (6)$$

where $f^p_{\mathbf{k}_i}$ and $f^p_{\mathbf{k}_j}$ are Fermi-Dirac population distributions for holes with momenta $\mathbf{k}_i$ and $\mathbf{k}_j$ respectively. The Cooper-pair electron term has to be calculated in a BCS state $I_e = \langle BCS|c^\dagger_{\mathbf{k}_1,J_1+\sigma_1}(t_1)c^\dagger_{\mathbf{k}_2,J_2+\sigma_2}(t_2)c_{\mathbf{k}_3,J_3+\sigma_3}(t_3)c_{\mathbf{k}_4,J_4+\sigma_4}(t_4)|BCS\rangle$ using the Bogoliubov transformation

$$c^\dagger_{\mathbf{k},J}(t) = e^{i\tilde{\mu}_n t}\left(u_\mathbf{k} e^{iE_\mathbf{k}t}\gamma^\dagger_{\mathbf{k},J} - s_J v_\mathbf{k} e^{-iE_\mathbf{k}t}\gamma_{-\mathbf{k},-J}\right) \qquad (7)$$

where $\gamma^\dagger_{\mathbf{k},J}$ is a Bogoliubov quasiparticle creation operator with crystal momentum $\mathbf{k}$ and angular momentum $J$, the electron energy above the quasi Fermi level, $E_{Fn}$, is $\xi_n(\mathbf{k}) = \mathbf{k}^2/2m_n - E_{Fn}$ and quasiparticle energy $E_\mathbf{k} = \sqrt{\xi_n^2(\mathbf{k})+\Delta^2}$, $\tilde{\mu}_n = E_{CB}+E_{Fn}$, $s_J = 1(-1)$ for $J = \uparrow(\downarrow)$, and $u_\mathbf{k} = \sqrt{1/2\left[1+\xi_n(\mathbf{k})/E_\mathbf{k}\right]}$ and $v_\mathbf{k} = \sqrt{1/2\left[1-\xi_n(\mathbf{k})/E_\mathbf{k}\right]}$. The Cooper-pair electron term is then:

$$I_e = u_{\mathbf{k}_1} v_{\mathbf{k}_1} u^*_{\mathbf{k}_3} v^*_{\mathbf{k}_3} \delta_{\mathbf{k}_1, J_1+\sigma_1, -\mathbf{k}_2, -(J_2+\sigma_2)} \delta_{-\mathbf{k}_3, -(J_3+\sigma_3), \mathbf{k}_4, J_4+\sigma_4} S_{J_1+\sigma_1} S_{J_3+\sigma_3} e^{i\tilde{\mu}_n t_1} e^{i\tilde{\mu}_n t_2} e^{-i\tilde{\mu}_n t_3} e^{-i\tilde{\mu}_n t_4} \times$$

$$\times \begin{pmatrix} e^{iE_{\mathbf{k}_1}(t_1-t_2)} e^{-iE_{\mathbf{k}_3}(t_3-t_4)} f^n_{\mathbf{k}_1}\left(1-f^n_{\mathbf{k}_3}\right) + e^{-iE_{\mathbf{k}_1}(t_1-t_2)} e^{iE_{\mathbf{k}_3}(t_3-t_4)} f^n_{\mathbf{k}_3}\left(1-f^n_{\mathbf{k}_1}\right) \\ -e^{-iE_{\mathbf{k}_1}(t_1-t_2)} e^{-iE_{\mathbf{k}_3}(t_3-t_4)}\left(1-f^n_{\mathbf{k}_1}\right)\left(1-f^n_{\mathbf{k}_3}\right) - e^{iE_{\mathbf{k}_1}(t_1-t_2)} e^{iE_{\mathbf{k}_3}(t_3-t_4)} f^n_{\mathbf{k}_1} f^n_{\mathbf{k}_3} \end{pmatrix} \quad (8)$$

where $f^n_{\mathbf{k}}$ is the Fermi-Dirac population distribution. For the HH band only, and assuming that $f^n_{\mathbf{k}}$ and $u_{\mathbf{k}}$ are slowly varying on the scale of $\mathbf{q}$, the density matrix is:

$$\rho(q_\mu, q_\nu)^{HH} = \sum_{\mathbf{k}} \pi \left|B_{\mathbf{k},\mathbf{q}_\mu}\right|^2 \left|B_{\mathbf{k},\mathbf{q}_\nu}\right|^2 \left|\Delta_{\mathbf{k}}/E_{\mathbf{k}}\right|^2 \left(f^p_{\mathbf{k}}\right)^2 \delta\left(\omega^{ph}_{\mathbf{q}_\nu} + \omega^{ph}_{\mathbf{q}_\mu} - 2\omega_{\mathbf{k}} - 2\tilde{\mu}_n\right) \times$$

$$\begin{pmatrix} 0 & 0 & 0 & 0 \\ 0 & 1 & 1 & 0 \\ 0 & 1 & 1 & 0 \\ 0 & 0 & 0 & 0 \end{pmatrix} \begin{pmatrix} \dfrac{f^n_{\mathbf{k}}(1-f^n_{\mathbf{k}})}{(E_{\mathbf{q}_\mu,\mathbf{k}}+E_{\mathbf{k}})^2} + \dfrac{f^n_{\mathbf{k}}(1-f^n_{\mathbf{k}})}{(E_{\mathbf{q}_\mu,\mathbf{k}}-E_{\mathbf{k}})^2} - \\ \dfrac{(1-f^n_{\mathbf{k}})(1-f^n_{\mathbf{k}})}{(E_{\mathbf{q}_\mu,\mathbf{k}}-E_{\mathbf{k}})(E_{\mathbf{q}_\mu,\mathbf{k}}+E_{\mathbf{k}})} - \dfrac{f^n_{\mathbf{k}} f^n_{\mathbf{k}}}{(E_{\mathbf{q}_\mu,\mathbf{k}}+E_{\mathbf{k}})(E_{\mathbf{q}_\mu,\mathbf{k}}-E_{\mathbf{k}})} + \\ + \dfrac{f^n_{\mathbf{k}}(1-f^n_{\mathbf{k}})}{(E_{\mathbf{q}_\nu,\mathbf{k}}+E_{\mathbf{k}})(E_{\mathbf{q}_\mu,\mathbf{k}}+E_{\mathbf{k}})} + \dfrac{f^n_{\mathbf{k}}(1-f^n_{\mathbf{k}})}{(E_{\mathbf{q}_\nu,\mathbf{k}}-E_{\mathbf{k}})(E_{\mathbf{q}_\mu,\mathbf{k}}-E_{\mathbf{k}})} - \\ \dfrac{(1-f^n_{\mathbf{k}})(1-f^n_{\mathbf{k}})}{(E_{\mathbf{q}_\nu,\mathbf{k}}-E_{\mathbf{k}})(E_{\mathbf{q}_\mu,\mathbf{k}}+E_{\mathbf{k}})} - \dfrac{f^n_{\mathbf{k}} f^n_{\mathbf{k}}}{(E_{\mathbf{q}_\nu,\mathbf{k}}+E_{\mathbf{k}})(E_{\mathbf{q}_\mu,\mathbf{k}}-E_{\mathbf{k}})} + (\mathbf{q}_\mu \longleftrightarrow \mathbf{q}_\nu) \end{pmatrix} \quad (9)$$

where $E_{\mathbf{q}_\mu,\mathbf{k}} = \omega^{ph}_{\mathbf{q}_\mu} - \omega_{\mathbf{k}} - \tilde{\mu}_n$, $E_{\mathbf{q}_\nu,\mathbf{k}} = \omega^{ph}_{\mathbf{q}_\nu} - \omega_{\mathbf{k}} - \tilde{\mu}_n$ and $(\mathbf{q}_\mu \longleftrightarrow \mathbf{q}_\nu)$ indicates another eight terms similar to the first eight, but with exchanged $q_\mu$ and $q_\nu$. The basis for the two-photon density matrix is circular right or left handed polarization $\left|R_{\mathbf{q}_\mu} R_{\mathbf{q}_\nu}\right\rangle, \left|R_{\mathbf{q}_\mu} L_{\mathbf{q}_\nu}\right\rangle, \left|L_{\mathbf{q}_\mu} R_{\mathbf{q}_\nu}\right\rangle, \left|L_{\mathbf{q}_\mu} L_{\mathbf{q}_\nu}\right\rangle$. The two-photon state is not only polarization-correlated, but is in fact a pure entangled state:

$$\left|\Psi_{ph}\right\rangle = \frac{1}{\sqrt{2}}\left(\left|R_{\mathbf{q}_\mu} L_{\mathbf{q}_\nu}\right\rangle + \left|L_{\mathbf{q}_\mu} R_{\mathbf{q}_\nu}\right\rangle\right) \quad (10)$$

This entangled state is a result of the correlated photon polarization with no which-path information; however the plus sign in Eq. 10 is different from the minus sign in the singlet state in Eq. 1. The reason for this difference is the effect of the many-body anti-symmetrization in the BCS state. This calculation result is also validated by the Feynman diagram approach. Each diagram in Fig. 2 b corresponds to one of the terms in the expression for the hole part of the calculation – $I_h$ (Eq. 6), and the negative sign between the first and second terms results from the fermionic exchange of holes (Fig. 2 b dashed lines). In the density matrix (Eq. 9), the non-vanishing diagonal elements are described by the first diagram in Fig. 2 b, and the off-diagonal elements by the second diagram. However the sign of the off-diagonal elements in Eq. 9 is changed due to the minus sign in the Bogoliubov transformation (Eq. 7), so that all four elements are positive. Conservation of angular momentum at each vertex of the diagrams then allows the determination of the photon polarization similar to the non-vanishing elements of the calculated density matrix (Eq. 9).

In a bulk semiconductor, the LH band adds other non-vanishing elements to the density matrix thus resulting in a mixed state (Fig. 3 a).

$$\rho(q_\mu, q_\nu) = \alpha \rho(q_\mu, q_\nu)^{HH} + \beta \rho(q_\mu, q_\nu)^{LH} \tag{11}$$

where $\rho(q_\mu, q_\nu)^{LH}$ is the contribution of the LH-CB transitions with non-vanishing elements corresponding to photons with identical circular polarization, and the coefficients $\alpha$ and $\beta$ depend on the population of the bands according to Eq. 9. The dipole moments of CB-LH transitions in direct-bandgap (e.g. Zincblende) materials are different from those of CB-HH [26], and therefore even for a bulk semiconductor with LH/HH degeneracy, the emitted two-photon state is not completely mixed. However, in a QW with a much smaller population of the LH band the purity of the state (Eq. 11) is enhanced (Fig. 3 b).

The purity of the generated entangled states, therefore, relies on the interplay between energy scales of the LH/HH splitting due to spin-orbit interaction (SOI), semiconductor bandgap, injected carrier density and the temperature. The largest energy scale in this scheme is the bandgap between the conduction $E_{CB}$ and the valence $E_{HH/LH}$ bands which determines the energy of the emitted photons. In typical AlGaAs structures the bandgap is larger than 1500meV [28] and is therefore much larger than all other energy scales in the system. The spin-orbit interaction in GaAs based materials is significant, but much smaller than the bandgap. In bulk GaAs, the split-off band is 340meV below the top valence band [29]. The SOI energy $\Delta E_{SOI}$ is thus smaller than the bandgap, but larger than other energies in our system. The superconducting gap, $\Delta$, of typical parent superconductors such as Nb is on the meV scale, while a proximity-induced gap in the semiconductor is even smaller [30], so that $E_{CB} - E_{HH/LH} \gg \Delta E_{SOI} \gg E_{HH} - E_{LH}, E_{Fp} - E_{LH}, E_{Fn} - E_{CB} \gg \Delta$. The injected electron and hole densities determine the locations of the conduction and the valence band quasi-Fermi levels relative to the band edges: $E_{Fn} - E_{CB}$, $E_{Fp} - E_{HH}$ and $E_{Fp} - E_{LH}$. At practical injection levels in typical optoelectronic materials such as AlGaAs this energy scale is around 10meV and can be calculated with analytical approximations [31]. The populations of the LH and the HH bands depend on the $\Delta E_H = E_{HH} - E_{LH}$ energy splitting determined by the thickness of the QW reaching values of tens of meV [29]. For a fixed carrier injection level and a given LH/HH splitting, the hole populations of the HH and the LH bands depend on temperature according to the Fermi-Dirac distribution $f^p(E)$ (Fig. 4a). High purity of the emitted entangled photons is obtained by reducing the LH contribution $\rho(q_\mu, q_\nu)^{LH}$ in Eq. 11, which can be obtained by lowering the temperature or by increasing the LH/HH splitting.

Typical LH/HH splitting in a QW can result in an entangled state with very high purity, whereas at smaller LH/HH splitting, $\Delta E_H$, the holes partially populate both HH and LH bands.

For lower temperature, T, the population of the LH band is smaller, and the CB-LH transition degrades the entangled state purity, given by $Tr\left[\rho(q_\mu,q_\nu)^2\right]$, less severely than at higher temperatures (Fig. 4 c). Nevertheless, even at higher temperatures, high purity of entanglement can be obtained by increasing the LH/HH separation (Fig. 4 b). In typical semiconductor QWs used in optoelectronics, the LH/HH separation of several tens of meV can be obtained for QW thickness smaller than 10nm [29]. Therefore mixing of the entangled states will be significant only close to room temperature. At temperatures below the superconducting transition of typical *s*-wave low-$T_c$ materials such as Nb [30], the emitted photons thus should be in an essentially pure entangled state. The superconducting proximity effect has been demonstrated recently with high-$T_c$ materials [ 32 ], enabling potential applications of this scheme at much higher temperatures in hybrid semiconductor-high-$T_c$ devices as well [33].

Another effect that could hinder the proper operation of the proposed device is strong disorder. Disorder-induced levels can lead to one-photon emission with energies above the bandgap $E_{BG}$ by normal electron-hole recombination. They can also lead to one-photon emission with energies below $E_{BG}$ – by recombination of a Cooper pair with a single hole, which results in a photon and an electron at a higher energy level. If the two-photon emission is selected at energies similar to those of the disorder-induced one-photon emission, the quality of the entanglement source can be affected. In order to ensure proper operation of the device, the energy difference of the photons in the pair must be larger than the energy broadening given by the disorder $\Delta E_D$, so that photon-pair emission can be spectrally filtered from disorder-induced one-photon emission. The energy broadening of the one-photon emission is limited by the level of the disorder in the system, whereas the desired energy difference between the photons in the entangled pair is limited by the separation between the light-hole and heavy-hole energies $\Delta E_H$ to ensure the generation of pure entangled states. In typical 10nm-thick AlGaAs QWs this separation is $\Delta E_H > 10 meV$ [29]. The most significant effect of disorder in QWs results from thickness variation, which leads to QW energy level variation. For this QW energy level variation to be comparable to $\Delta E_H$, the variation in thickness of more than 100% is required [34]. In modern high-quality AlGaAs QWs, the thickness variation can be controlled on a monolayer level, resulting in very narrow linewidths of around $\Delta E_D \sim 0.5 meV$ which ensures the desired condition: $\Delta E_D << \Delta E_H$. This narrow linewidth also enables strongly-coupled light-matter interaction in microcavities and was measured both spectrally [35], and with ultrafast pump-probe experiments [36].

Two-photon emission in our scheme can be spectrally selected at energies, different by more than 10meV. For the disorder-induced one photon emission also to be at energies different by 10meV, that must be the energy scale of the broadening $\Delta E_D$ introduced by disorder with the corresponding relaxation time $\tau \sim 60 fs$. This short relaxation time corresponds to a very short mean free path of $l_f = v_F \tau < 10 nm$, where $v_F$ is the Fermi velocity. As the second-order Cooper pair recombination rate has been shown to be comparable to the first-order normal recombination rate [21-23], this estimation provides the requirements on the amount of disorder for proper operation of the entanglement sources. Moreover, if we perform an extremely pessimistic estimation where we completely neglect the enhancement of recombination by superconductivity, the estimation does not change significantly. In a hypothetical scenario without superconducting enhancement, the rate of one photon emission in semiconductors at a given energy is typically about 6 orders of magnitude stronger than the rate of two-photon emission [8]. For a typical Gaussian distribution of disorder-induced energy levels in epitaxial QWs [37], the rate of one-photon emission will be weaker than the rate of two-photon emission (without superconducting enhancement) at energies above $5\sigma$ in the distribution because of the low density of disorder-induced states above $5\sigma$. If the $5\sigma$ energy is 10meV (determined by $\Delta E_H$), the disorder-induced energy width, $\Delta E_D$, has to be below ~4meV, corresponding to a mean-free path longer than 25nm, for proper operation of the source. This estimated mean free path is extremely short − almost two orders of magnitude shorter than the typical values in AlGaAs QWs [35, 36], and about three orders of magnitude shorter than in high-quality AlGaAs heterostructures [38]. Therefore, for devices based on typical QWs, disorder-induced one-photon emission does not affect the operation of the proposed device, and for low-quality structures, the disorder-induced mean free path should be longer than the critical length of about 10nm (25nm if superconducting enhancement [21-23] is neglected ).

In conclusion, we have shown that polarization-entangled photons can be generated efficiently by Cooper-pair luminescence in semiconductors without isolated emitters. Due to the lack of which-path information in the second-order transition and the lifted degeneracy of the valence bands in QWs, the emission results in pure-state polarization-entangled photons, with generation rates enhanced by the macroscopic coherence of the superconducting state. The proposed source of entangled photons can provide insights into the physics of superconductivity and light-matter interaction in solids, as well as enable practical applications in quantum technologies.

We appreciate financial support from the Natural Sciences and Engineering Research Council of Canada (NSERC) and from the Canadian Institute for Advanced Research (CIFAR).

## References:

[1] G. Weihs, T. Jennewein, C. Simon, H. Weinfurter and A. Zeilinger, "Violation of Bell's inequality under strict Einstein locality conditions", Phys. Rev. Lett. **81**, 5039 (1998).

[2] P. G. Kwiat, K. Mattle, H. Weinfurter, A. Zeilinger, A. V. Sergienko and Y. Shih, "New High-Intensity Source of Polarization-Entangled Photon Pairs", Phys. Rev. Lett. **75**, 4337 (1995).

[3] N. Gisin G. Ribordy, W.Tittel, and H. Zbinden, " Quantum cryptography" , Rev. Mod. Phys. **74**, 145 (2002); T. Jennewein, C. Simon, G. Weihs, H. Weinfurter, and A. Zeilinger, "Quantum cryptography with entangled photons," Phys. Rev. Lett. **84**, 4729 (2000).

[4] J. L. O'Brien, "Optical quantum computing," Science 318, 1567 (2007).; T. D. Ladd, F. Jelezko, R. Laflamme, Y. Nakamura, C. Monroe, J. L. O'Brien, "Quantum Computers", Nature **464**, 45 (2010).

[5] T. Nagata, R. Okamoto, J. L. O'Brien, K. Sasaki, S. Takeuchi, "Beating the Standard Quantum Limit with Four-Entangled Photons", Science, **316**, 726 (2007); M.W. Mitchell, J.S. Lundeen, and A.M. Steinberg, "Super-resolving phase measurements with a multi-photon entangled state," Nature **429**, 161 (2004).

[6] A. G. White, D. F. V. James, P. H. Eberhard, and P. G. Kwiat, "Nonmaximally entangled states: production, characterization, and utilization", Phys. Rev. Lett., **83**, 3103 (1999).

[7] R. W. Boyd, "Nonlinear optics", third edition, Academic press, London, U.K. (2008).

[ 8 ] A. Hayat, P. Ginzburg, M. Orenstein, "Observation of two-photon emission from semiconductors," Nature Photon. **2**, 238 (2008); A. Hayat, P. Ginzburg and M. Orenstein "Measurement and model of the infrared two-photon emission spectrum of GaAs", Phys. Rev. Lett., **103**, 023601 (2009).

[9] Y. Ota, S. Iwamoto, N. Kumagai, and Y. Arakawa, "Spontaneous Two-Photon Emission from a Single Quantum Dot", Phys. Rev. Lett. **107**, 233602 (2011).

[10]A. N. Poddubny, P. Ginzburg, P. A. Belov, A. V. Zayats, and Y. S. Kivshar, "Tailoring and enhancing spontaneous two-photon emission using resonant plasmonic nanostructures", Phys.

Rev. A **86**, 033826 (2012); A. Nevet, N. Berkovitch, A. Hayat, P. Ginzburg, S. Ginzach, O. Sorias, M. Orenstein, "Plasmonic nano-antennas for broadband enhancement of two-photon emission from semiconductors", Nano Letters, **10**, 1848 (2010).

[11] R. M. Stevenson, R. J. Young, P. Atkinson, K. Cooper, D. A. Ritchie, A. J. Shields, "A semiconductor source of triggered entangled photon pairs", Nature 439, 179 (2006); Akopian, N. H. Linder, E. Poem, Y. Berlatzky, J. Avron, D. Gershoni, B. D. Geradot, and P. M. Petroff," Entangled Photon Pairs from Semiconductor Quantum Dots", Phys. Rev. Lett. **96**, 130501 (2006).

[12] E. A. Meirom, N. H. Lindner, Y. Berlatzky, E. Poem, N. Akopian, J. E. Avron, and D. Gershoni, "Distilling entanglement from random cascades with partial "which path" ambiguity", Phys. Rev. A **77**, 062310 (2008).

[13] S. De Franceschi, L. Kouwenhoven, C. Schönenberger and W. Wernsdorfer, "Hybrid superconductor–quantum dot devices", Nature Nanotech. **5**, 703 (2010).

[14] G. Katsaros, P. Spathis, M. Stoffel, F. Fournel, M. Mongillo, V. Bouchiat, F. Lefloch, A. Rastelli, O. G. Schmidt and S. De Franceschi, "Hybrid superconductor–semiconductor devices made from self-assembled SiGe nanocrystals on silicon", Nature Nanotech. **5**, 458 (2010).

[15] V. Mourik, K. Zuo, S. M. Frolov, S. R. Plissard, E. P. A. M. Bakkers, L. P. Kouwenhoven, "Signatures of Majorana fermions in hybrid superconductor-semiconductor nanowire devices", Science, **336**, 1003 (2012).

[16] P. Recher, Y. V. Nazarov, and L. P. Kouwenhoven, "Josephson Light-Emitting Diode", Phys. Rev. Lett. **104**, 156802 (2010).

[17] F. Godschalk, F. Hassler, and Y. V. Nazarov, "Proposal for an Optical Laser Producing Light at Half the Josephson Frequency", Phys. Rev. Lett. **107**, 073901 (2011).

[18] I. Suemune, T Akazaki, K. Tanaka, M. Jo, K. Uesugi, M. Endo, H. Kumano, E. Hanamura, H. Takayanagi, M. Yamanishi and H. Kan, "Superconductor-Based Quantum-Dot Light-Emitting Diodes: Role of Cooper Pairs in Generating Entangled Photon Pairs", Jpn. J. Appl. Phys. **45**, 9264 (2006).

[19] F. Hassler, Y. V. Nazarov, L. P. Kouwenhoven, "Quantum manipulation in a Josephson light-emitting diode", Nanotechnology, **21**, 274004 (2010).

[20] M. Khoshnegar and A. H. Majedi, "Entangled photon pair generation in hybrid superconductor–semiconductor quantum dot devices", Phys. Rev. B **84**, 104504 (2011).

[21] H. Sasakura, S. Kuramitsu, Y. Hayashi, K. Tanaka, T. Akazaki, E. Hanamura, R. Inoue, H. Takayanagi, Y. Asano, C. Hermannstädter, H. Kumano, and I. Suemune, "Enhanced Photon Generation in a Nb/n-InGaAs/p-InP Superconductor/Semiconductor-Diode Light Emitting Device", Phys. Rev. Lett. **107**, 157403 (2011).

[22] I. Suemune, Y. Hayashi, S. Kuramitsu, K. Tanaka, T. Akazaki, H. Sasakura, R. Inoue, H. Takayanagi, Y. Asano, E. Hanamura, S. Odashima, and H. Kumano, "A Cooper-Pair Light-Emitting Diode: Temperature Dependence of Both Quantum Efficiency and Radiative Recombination Lifetime", Appl. Phys. Express **3**, 054001 (2010).

[23] Y. Asano, I. Suemune, H. Takayanagi, and E. Hanamura, "Luminescence of a Cooper Pair", Phys. Rev. Lett. **103**, 187001 (2009).

[24] M. Tinkham, "Introduction to Superconductivity", McGraw-Hill, New York, ed. 2, (1996).

[25] A. Kastalsky, A. W. Kleinsasser, L. H. Greene, R. Bhat, F. P. Milliken, and J. P. Harbison, "Observation of pair currents in superconductor-semiconductor contacts", Phys. Rev. Lett. **67**, 3026–3029 (1991).

[26] C. C. Lee and H. Y. Fan., "Two-photon absorption with exciton effect for degenerate valence bands", Phys. Rev. B **9**, 3502–3516 (1974).

[27] A. A. Abrikosov, L. P. Gorkov, I. E. Dzyaloshinski, "Methods of quantum field theory in statistical physics", Prentice Hall, Englewood Cliffs, N.J (1963).

[28] A.K. Saxena, "The conduction band structure and deep levels in $Ga_{1-x}Al_xAs$ alloys from a high-pressure experiment", J. Phys. C., **13**, 4323-4334 (1980).

[29] G. Bastard, "Wave Mechanics Applied to Semiconductor Heterostructures." Wiley-Interscience. (1991).

[30] D. R. Heslinga, S. E. Shafranjuk, H. van Kempen T. M. Klapwijk, "Observation of double-gap-edge Andreev reflection at Si/Nb interfaces by point-contact spectroscopy", Phys. Rev. B **49**, 10484–10494 (1994).

[31] W. B. Joyce and R. W. Dixon, "Analytic approximations for the Fermi energy of an ideal Fermi gas", Appl. Phys. Lett. **31**, 354 (1977).

[32] P. Zareapour, A. Hayat, S.Y.F. Zhao, M. Kreshchuk, A. Jain, D. C. Kwok, N. Lee, S.-W. Cheong, Z. Xu, A. Yang, G. D. Gu, R. J. Cava, K.S. Burch, "Proximity-induced high-temperature superconductivity in topological insulators $Bi_2Se_3$ and $Bi_2Te_3$. ", Nature Commun. **3**, 1056 (2012).

[33] A. Hayat, P. Zareapour, S.Y.F. Zhao, A.Jain, I.G. Savelyev, M. Blumin, Z. Xu, A.Yang, G. D. Gu, H. E. Ruda, S. Jia, R. J. Cava, A. M. Steinberg, and K. S. Burch, "Hybrid high-temperature superconductor-semiconductor tunnel diode." Phys. Rev. X **2**, 041019 (2012).

[34] R. L. Greene, K. K. Bajaj and D. E. Phelps, "Energy levels of Wannier excitons in GaAs-Ga1-xAlxAs quantum-well structures", Phys. Rev. B **29**, 1807–1812 (1984).

[35] H. M. Gibbs, G. Khitrova and S. W. Koch, "Exciton–polariton light–semiconductor coupling effects", Nature Photon. 5, 273 (2011).

[36] A. Hayat, C. Lange, L. A. Rozema, A. Darabi, H. M. van Driel, A. M. Steinberg, B. Nelsen, D. W. Snoke, L. N. Pfeiffer, K.W. West, "Dynamic Stark effect in strongly coupled microcavity exciton-polaritons ", Phys. Rev. Lett. **109**, 033605 (2012).

[37] M. A. Herman, D. Bimberg, and J. Christen., "Heterointerfaces in quantum wells and epitaxial growth processes: Evaluation by luminescence techniques", J. Appl. Phys. **70**, R1 (1991); D. Bimberg, D. Mars, J. N. Miller, R. Bauer, and D. Oertel, "Structural changes of the interface, enhanced interface incorporation of acceptors, and luminescence efficiency degradation in GaAs quantum wells grown by molecular beam epitaxy upon growth interruption", J. Vac. Sci. Technol. B **4**, 1014 (1986).

[38] M. Avinun-Kalish, M. Heiblum, O. Zarchin, D. Mahalu and V. Umansky, "Crossover from 'mesoscopic' to 'universal' phase for electron transmission in quantum dots", Nature **436**, 529 (2005).

**Figure captions:**

**Figure 1.** Energy level diagram of Cooper-pair luminescence in a direct-bandgap semiconductor in the (incorrect) one-particle picture: (a) bulk (b) QW. (c) Energy level diagram of Cooper-pair luminescence in a QW in the correct two-particle picture.

**Figure 2** (a) Feynman diagrams of the emission process. The wavy lines indicate photons, dashed lines indicate holes, and the solid lines indicates electrons. The double-arrowed electron propagators describe the Green functions resulting from non-vanishing $\langle BCS | c^\dagger_{\mathbf{k}_1,J_1} c^\dagger_{\mathbf{k}_2,J_2} | BCS \rangle$ terms. (b) Feynman diagrams of the density matrix calculations.

**Figure 3.** Calculated density matrix of the two-photon polarization state for Cooper-pair luminescence. (a) in a bulk direct-bandgap semiconductor (b) in a QW with large LH/HH splitting.

**Figure 4.** (a) A schematic diagram of the energy scales. The solid lines indicate band edge energies, the dashed lines indicate quasi-Fermi levels; the dotted curves indicate the Fermi-Dirac population distributions. (b,c) Calculated purity of the photon polarization-entangled state (b) vs. LH-HH energy splitting for different temperatures (c) vs. temperature for different LH-HH energy splitting, where $E_{LH}$ and $E_{HH}$ are LH and HH band edge energies respectively, $E_{Fp}$ is the valence band quasi-Fermi level, and $T$ is the temperature.

# Figure 1

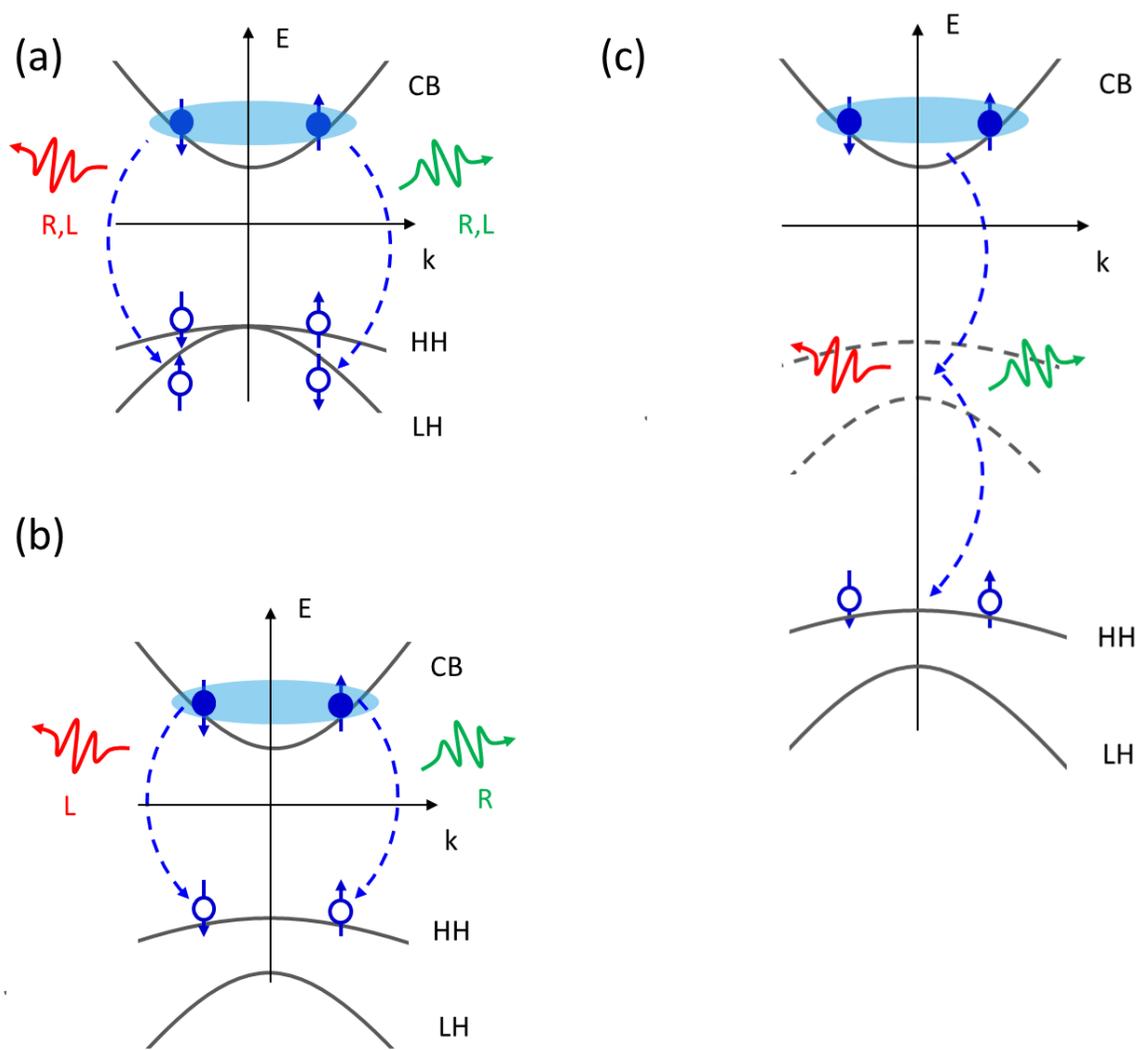

**Figure 2**

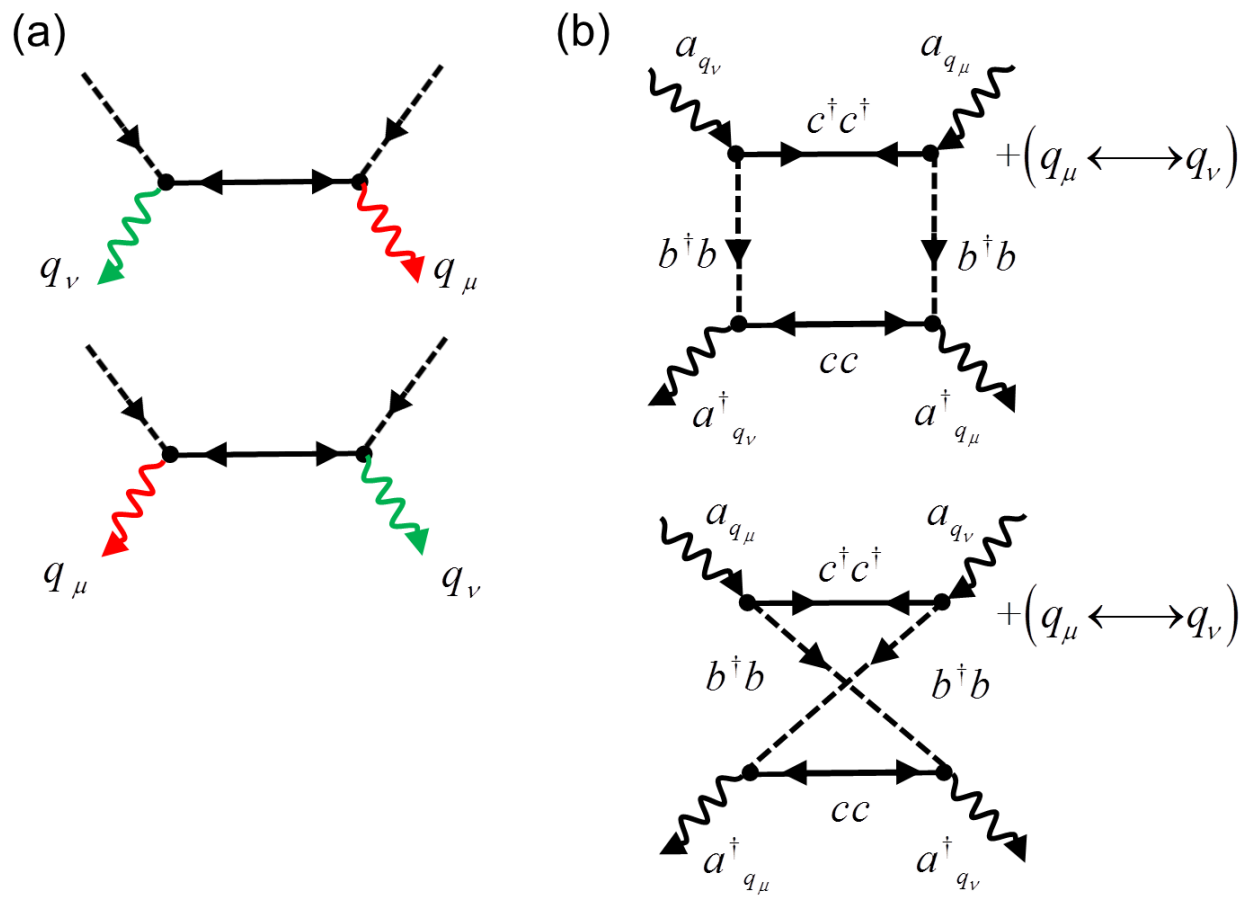

**Figure 3**

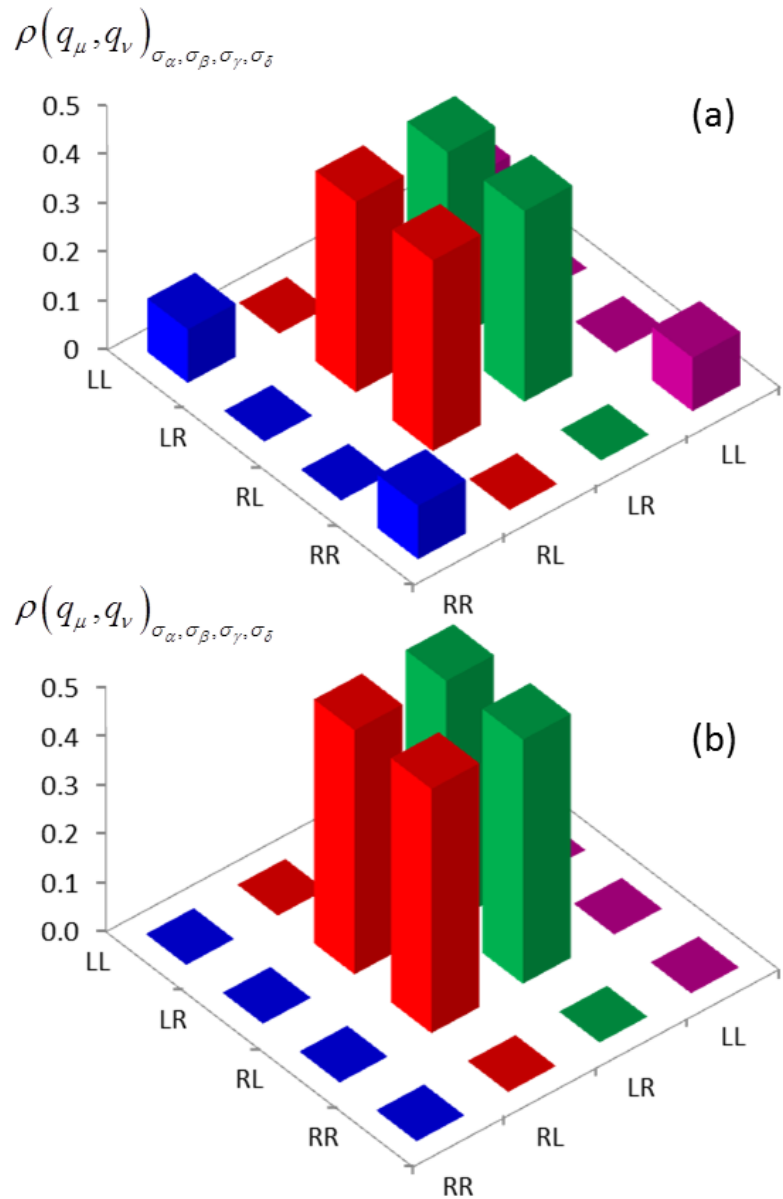

**Figure 4**

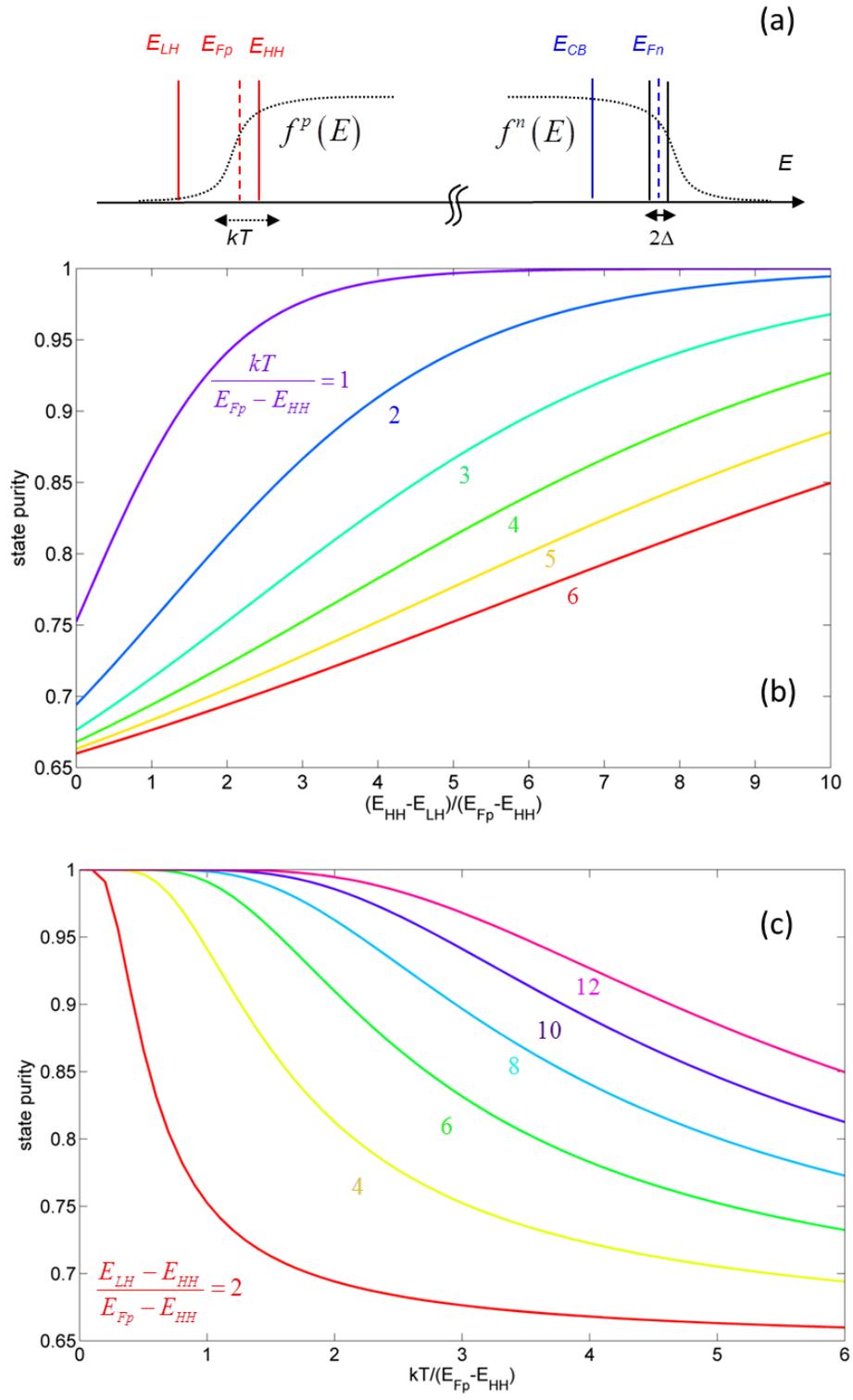